\documentclass[]{raa_rb}            % referee version: for submission
\usepackage{graphicx,times}
\usepackage{natbib}

\providecommand{\bm}[1]{\mbox{\boldmath$#1$\unboldmath}}

\begin{document}
\title{Statistical properties of the Bipolar Magnetic Regions
%\footnotetext{\small $*$ Supported by the National Natural Science Foundation of China.}
}

\volnopage{ {\bf --} Vol.\ {\bf --} No. {\bf XX}, 000--000}
   \setcounter{page}{1}

\author{Dong Li
      \inst{1}
   }
%% Here is an example of three authors come from different institutes.
%% For single author or all the authors from an institute, use "\inst{}" only

\institute{Key Laboratory of Dark Matter and Space Science, Purple
Mountain Observatory, Chinese Academy of Sciences, Nanjing 210008,
China; {\it lidong@pmo.ac.cn}
%% Please give the E-mail address of the author, to whom future correspondence and
%% offprint requests will be sent.
%
%
%
%\vs \no
%   {\small Received [year] [month] [day]; accepted [year] [month] [day] }
}

\abstract{Using the observations from Michelson Doppler Imager (MDI)
onboard Solar and Heliospheric Observatory (SOHO), we develop a
computational algorithm to automatically identify the bipolar
magnetic regions (BMRs) in the active regions, and then study their
statistical properties. The individual magnetic (positive or
negative) pole of the BMR is determined from the region with an
absolute strength above 55 G and with an area above 250 pixel$^2$
($\sim$495 Mm$^2$), while a BMR is identified as a pair of positive
and negative poles with a shortest area-weight distance between
them. Based on this method, 2234 BMRs are identified from the MDI
synoptic magnetograms between the Carrington Rotation 1909 (1996 May
06) and 2104 (2010 December 10). 1005 of them are located in the
northern hemisphere, while the other 1229 are in the southern
hemisphere. We find that the BMR paraments (e.g., latitudes,
separations, fragments, and strength) are similar to those of active
regions (ARs). Moreover, based on the maximum likelihood estimation
(MLE) method, the frequency distributions of these BMRs occurrence
as functions of area size and magnetic flux exhibit a power-law
behavior, i.e., $dN/dx \propto {\bf x^{-\alpha_{x}}}$, with an index
of $\alpha_{A}$ = 1.98 $\pm$ 0.06 and $\alpha_{F}$ = 1.93 $\pm$
0.05. We also find that their orientation angles ($\theta$) follow
the ``Hale's Polarity Laws" and deviate slightly to the solar
equator direction. Consistent with the previous findings, we obtain
the orientation angles dependence on the latitudes for the normal
BMRs during the 23rd solar cycle. The north-south asymmetry of these
BMRs is also detected here. \keywords{methods: statistical; Sun:
activity; Sun: magnetic fields} }

\authorrunning{Dong Li}            %author_head in even pages
\titlerunning{Statistical properties of the Bipolar Magnetic Regions}  % title_head in odd pages
\maketitle

%% The author head (on even pages) and the title head (on odd pages) will be
%% automatically extracted from \author{} and \title{}. Whenever the title is too long,
%% you will be asked to supply a shorter one by inserting either \authorrunning{} or
%% \titlerunning{} before \maketitle. Anyway, you can specify your own heads.
%%
%%
%% Note: In the following text body of your manuscript, please note several differences from
%%       other major journals:
%% (1) \subsection{Please Capitalize the First Letter of Each Notional Word in Subsection Title}
%% (2) Please Capitalize the First Letter of Each Notional Word in all tables' captions

%________________________________________________ sections below

\section{Introduction}           %% first-level sections will be auto-capitalized
\label{intro} The magnetic fields are believed as the dominant
reasons for the evolution of solar activity, while the stronger and
larger magnetic fields on the solar surface are mainly in the active
regions (ARs). Meanwhile, most of the energetic and geo-effective
events take place at the ARs, such as solar flares, coronal mass
ejections (CMEs), solar energetic particle events, and eruptive
prominences. Therefore, quantitative study the magnetic fields in
the ARs is important to the basic solar physics. As earlier as more
than 350 years ago, ARs have been studied as sunspots on white light
images
\citep[e.g.,][]{Wolf61,Maunder04,McKinnon87,Hathaway99,Li01,Hathaway03,Zhang10,Jiang11,Hathaway10}.
It is well known that the number of the sunspots on solar disk
display a periodic behavior, which has an average period of about 11
years. And the positions of the sunspots exhibit the butterfly
shapes, which is well-known as ``Butterfly Diagram", this suggests
that the behavior of the sunspot follows the ``Sp\"{o}rer's Law of
Zones". The sunspots are those regions with stronger magnetic fields
on the Sun, and their magnetic nature is followed by the famous
``Hale's Polarity Laws''. Both the unipolar spots and the preceding
members of the bipolar spots, their magnetic polarity is negative
before the last sunspot minimum and positive since the solar minimum
in the northern hemisphere, while in the southern hemisphere, their
magnetic polarity is positive before the last sunspot minimum and
negative since the solar minimum \citep{Hale19}. That is to say, the
signs of the sunspot could be reversed at the solar minimum, thus
the period of the solar magnetic fields is about 22 years, which is
called as solar magnetic cycle.

The characteristics of the magnetic fields at ARs have been reported
by many authors \citep[e.g.,][]{Howard89,Wang89,Harvey93,Zhang10}.
Using the Mount Wilson daily magnetogram data set, \cite{Howard89}
presented various properties of the magnetic fields at solar ARs.
For example, the average separation of the magnetic polarity was
about 7 deg ($\sim$86 Mm), the distribution of the magnetic flux per
AR showed a peak of about 2 $\times$ 10$^{21}$ Mx. This was similar
to the value of 4 $\times$ 10$^{21}$ Mx obtained by \cite{Wang89},
who used the data from National Solar Observatory/Kitt Peak during
1976$-$1986. Latter, \cite{Harvey93} studied the properties of the
ARs from the NSO/KP full-disk magnetogram during1975$-$1986
throughout Solar Cycle 21, and their conclusion was that the shape
of the characteristic size distribution for the ARs was a
fundamental invariant property of solar magnetic activity. Recently,
using the high-resolution synoptic magnetograms constructed from
SOHO/MDI image during 1996$-$2008, \cite{Zhang10} identified 1730
ARs and quantified their physical properties. The mean and maximum
magnetic flux of the individual ARs were 1.67$\times$10$^{22}$ Mx
and 1.97$\times$10$^{23}$ Mx, while those of each Carrington
rotation were 1.83$\times$10$^{23}$ Mx and 6.96$\times$10$^{23}$ Mx,
respectively.

A number of literatures
\citep[e.g.,][]{Bogdan88,Abramenko05,Canfield07,Zhang10} have
reported the distribution of magnetic flux or area of ARs and
sunspots. These papers reported that the distribution related to ARs
are usually log-normal. For example, \cite{Zhang10} have analyzed
the frequency distribution of ARs, and found that the distribution
with the function of area size and magnetic flux following a
log-normal function, this was consistent with the results obtained
by \cite{Abramenko05} and \cite{Canfield07}. And \cite{Bogdan88}
found the sunspot umbral area were distributed lognormally with
Mount Wilson white-light data in the interval between 1917 and 1982.
However, \cite{Parnell09} analyzed the magnetograms from SOHO/MDI
and Hinode/SOT, they found that all feature fluxes following the
power law distribution with a slope of 1.85 $\pm$ 0.14. While
\cite{Tang84} studied the 15 years of ARs data using the Mount
Wilson daily magnetograms data from 1967 to 1981, and their results
revealed that the number of the ARs decreased exponentially with the
increasing AR sizes. This was also proved by \cite{Zharkov05}, who
found that the number of sunspots growing nearly exponentially with
their area decreasing. Meanwhile, \cite{Harvey93} have successfully
used the polynomial function to fit the NSO/KP data. These different
fit functions are caused by the different observation data, and they
may be related to the physical mechanism of the ARs emergences. The
log-normal distribution has been regarded as the result of magnetic
fragmentation in the solar envelope \citep{Bogdan88}. While the
log-normal distribution of the AR's flux may also suggest that the
process of fragmentation dominates over the process of concentration
in the formation of the magnetic structure in ARs
\citep{Abramenko05,Canfield07}. On the other hand, the power-law
distribution of the AR's flux possibly suggests a self-similar
nature of all ARs \citep{Mcateer05}. If considering the magnetic
flux between 2$\times$10$^{17}$ Mx and 10$^{23}$ Mx, then the
power-law distribution also suggests that the mechanisms of the
surface magnetic features are scale-free \citep{Parnell09}. Finally,
\cite{Schrijver97} thought that the exponential distribution was
lead by the frequent fragmentation and collision (or merging) of the
magnetic features.

The orientation angles of the magnetic fields at ARs are important
to understand the solar physics, they not only represent the
important quantity that is related to the large-scale properties of
magnetic field distribution, but may be also related to the dynamo
progress which is believed to lead the solar activity cycle
\citep{Babcock61,Leighton64,Leighton69,Sheeley85}. \cite{Howard89}
have studied the orientation angles of the magnetic regions at ARs
with Mount Wilson magnetograms. He found that the distribution of
ARs orientation angles showing two broad maxima centered on the
`normal' orientation, and the reversed oriented ARs tending to be
relatively evenly distributed in orientation angle compared to the
normally oriented ones. However, the most studies are the tilt
angles of the magnetic regions, which are similar to the orientation
angles but not the same. The orientation angle is defined as the
angle from the positive/negative towards negative/positive fields
\citep{Howard89}. While the tilt angle is defined to be the angle
between the bipolar axis line and heliographic east-west line
\citep{Wang89}, it is the angle measured positive for magnetic
regions with leading fields equatorward of following fields and
negative for magnetic regions with leading fields poleward of
following fields \citep{Howard91a,Lij12}. In other word, the tilt
angle is very useful to study the Hale's and Joy's laws
\citep{Lij12}, while the orientation angle only exhibits the incline
of the ARs on solar disk. The first detailed study the tilt angles
of the magnetic regions is \cite{Hale19}, who found that the average
tilt angles of sunspots increasing with solar latitudes. This result
was confirmed by \cite{Brunner30}, who also found that the larger,
well-developed sunspots tending to have small tilt angles than the
smaller, less-developed ones. Then \cite{Wang89} used the NSO
magnetograms to find that the average tilt angles of all BMRs
relating to the east-west line showed a progressive increasing
toward high latitude, and the value was close to 9$^{\circ}$. After
that, \cite{Howard91a} studied the tilt angles of the magnetic
regions in ARs with Mount Wilson magnetograms, he found that the
variation of tilt angles with solar latitudes was not dependent on
the solar cycle phases. These ARs with the larger absolute tilt
angles have the rapid separation of the magnetic poles on average,
and their sizes are smaller than those with smaller absolute tilt
angles. Lately, \cite{Howard91b} studied the tilt angles of the
sunspots with Mount Wilson daily white-light photographs, he
obtained the similar results to the earlier studies about ARs
\citep[e.g.,][]{Wang89,Howard91a}, and the average tilt angle of the
sunspots was 4.2 $\pm$ 0.2 degree in his study.

In this paper, we automatically identify the BMRs in the ARs with
the high-resolution Carrington rotation synoptic magnetogram charts
constructed from SOHO/MDI observations between 1996 and 2010, and
then study their orientation angles. This paper is organized as
following: the observation and data reduction are introduced in
Section~\ref{Obs}, and the observation results are given in
Section~\ref{Res}, then our conclusions and discussions are given in
Section~\ref{Con}.

\section{Observations and Data reduction}
\label{Obs}

\subsection{Observations}
The data used here are from the SOHO/MDI magnetograms
\citep{Scherrer95}. MDI is designed to measure the velocity,
intensity and magnetic fields in the photosphere, and further to
study the magnetic fields in the corona. It can be detected the full
solar disk magnetogram with the spatial resolution of
$\sim$2$^{\prime\prime}$ pixel$^{-1}$ in every 96 minutes
\citep{Domingo95}. However, the full disk MDI magnetograms are not
used directly in this paper, but the Carrington Rotation (CR)
synoptic charts are used to analyze, and they can be downloaded from
the MDI homepages. The CR synoptic charts from MDI have two forms:
magnetic field and intensity synoptic charts. Only the magnetic
field charts are used in this paper. They are generated from the MDI
magnetograms at level 1.8. They are well re-calibrated, and several
observations of every location which have been collected over the
course of a solar rotation ($\sim$27 days) are averaged to make up
the charts. Therefore, the strength of the magnetic field at each
synoptic grid point is averaged, and they have been corrected by the
differential rotation before. Through the averaging process, the
effects of cosmic rays have also been reduced. The projection effect
of the magnetic filed have been corrected by assuming that MDI makes
line-of-sight measurements of a radial magnetic field. Finally, the
correction to the pixel area has also been performed in this paper
with a scaling factor \citep[see.,][]{Berger03,Tran05,Ulrich09}.

The finally synoptic charts of the magnetic fields have two versions
of the maps: a radial and a line-of-sight (LOS) version. The
synoptic charts about the projection effect along the longitude are
much better than the snapshot magnetograms, but they have lost the
temporal resolution. The noise level of the synoptic charts is about
5 Gauss, and the resolution of the synoptic charts have been changed
to a 3600 $\times$ 1080 pixel synoptic map. The carrington longitude
is linear, while the carrington latitude is sine. Noting that the
data at each longitude in these synoptic charts is observed at
different time, so the longitude in the synoptic charts is also
represent the time information.

\begin{figure}[h!!!]
   \centering
   \includegraphics[width=13.0cm, angle=0]{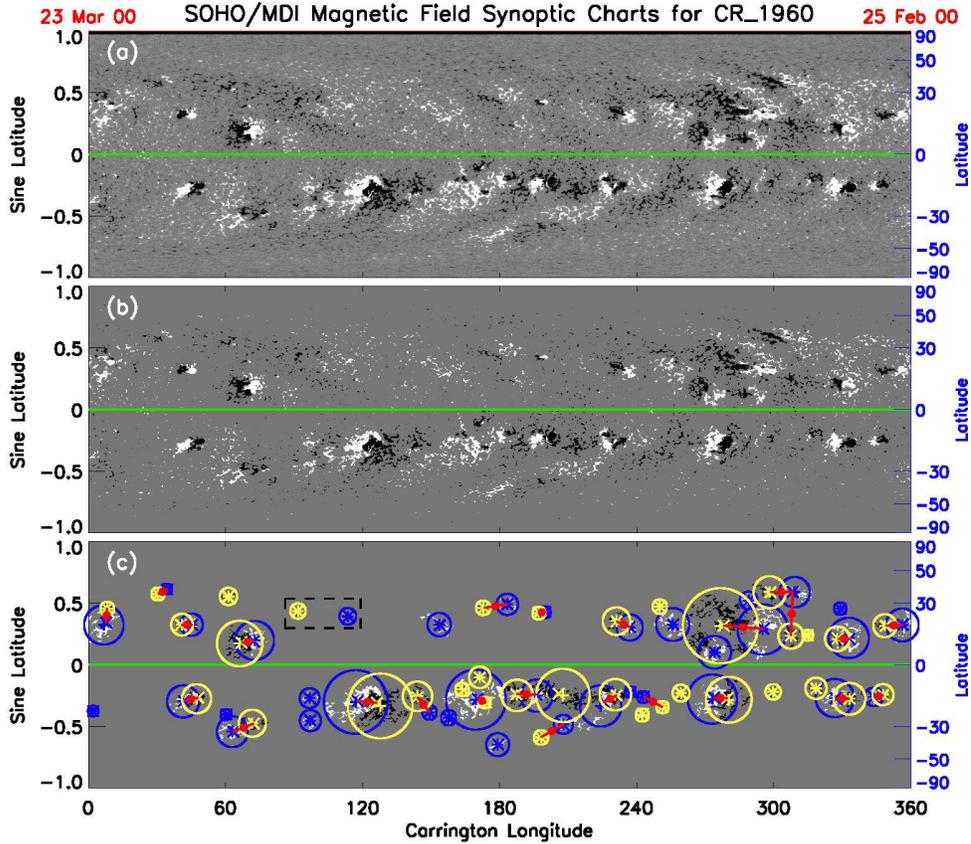}
   \caption{(a): The LOS magnetic field Synoptic Charts for Carrington
Rotation 1960 near central meridian, which started on 25 February
2000 and ended on 23 March 2000. (b): The magnetogram after
subtracting the strength threshold from panel (a); (c): The
magnetogram after subtracting the area threshold from panel (b), the
blue and yellow circles are the positive and negative poles, while
the `+' and `$\times$' represent the geometric centers and the
flux-weighted centers, respectively; the identified BMRs are
connected with the red lines, the filled circles are the positions
of the BMRs. The green line in each panel represents the solar
equator.}
   \label{fig1}
\end{figure}

All the synoptic maps, whether the radial or the LOS are constructed
from the data which observed by the disk meridians. These maps used
the data that observed near central meridian (0$^{\circ}$) are
called ordinarily synoptic charts, and additional charts are using
the data from other disk meridians, such as 60$^{\circ}$E,
45$^{\circ}$E, 30$^{\circ}$E, 15$^{\circ}$E, 15$^{\circ}$W,
30$^{\circ}$W, 45$^{\circ}$W, and 60$^{\circ}$W. However, those
additional charts have the disadvantage of disk longitude
offsetting, the ordinarily synoptic charts have not, so only the
ordinarily synoptic charts are used in this paper. Therefore, only
the LOS magnetic field observed near central meridian for Carrington
Rotation synoptic charts are used to identified the BMRs.
Figure~\ref{fig1} (a) shows an example of the synoptic images which
we use in this paper.

\subsection{Data reduction}
Two magnetic poles of the BMRs in the active regions are defined by
the threshold method, which include the strength and area
thresholds. As illustrated in Figure~\ref{fig1}, there are
essentially two steps, each of which corresponds to one panel in
this figure. The first step is to define the magnetic poles of the
BMRs. The positive and negative magnetite fields are separated from
the observation data, and the two magnetic poles are identified
respectively. Then we determine the strength threshold of the
positive or negative magnetic filed for every magnetogram with
equation (1). However, the magnetograms used in this paper are from
May 1996 to December 2010, which included the whole 23rd solar cycle
and the beginning phase of 24rd solar cycle. The strength thresholds
with equation~\ref{eq1} for different phases of the magnetograms
have large difference. To rule out this difference, we take the
average value ($\overline{TH}$) of all thresholds, and this value is
55 G for the positive filed and -55 G for the negative field. This
value is similar to that used by \cite{Zhang10}, who have been
identified the ARs by the minimum magnetic filed of 50 G with
SOHO/MDI synoptic magnetograms. Figure~\ref{fig1} (b) gives the
results after subtracting the $\overline{TH}$ from the original
observation data.

\begin{equation}
\label{eq1}
    TH = \mu \pm \gamma\cdot\sigma,
\end{equation}
In equation (1), $\mu$ is the mean value, and $\sigma$ represents
the standard deviation, while $\gamma$ is a constant which
determines empirically based on the type features to be detected. In
this paper we set $\gamma$ = 2, which is same as \cite{Colak08} to
determine the magnetic polarities of ARs. While `+' is for the
positive filed and `-' is for the negative filed.

Next, we will identify the two poles of the BMRs by the area
threshold. The magnetic images have been separated several
individual unconnected regions after subtracting the strength
threshold, and these regions can be marked automatically by the code
of LABEL\_REGION.pro in Interactive Data Language (IDL). However,
these isolated regions are not considered as the magnetic poles, for
these isolated regions which close to each other may be the same
magnetic pole. Therefore, we have to determine the area of the
magnetic poles. That is to say, it is possible that these closer
regions belong to the same magnetic pole. So the positions
(geometric centers) and the equivalent diameters (consider the
isolated region as a circle) of these isolated regions have been
calculated; and for every isolated region, we regard the geometric
center as the reference point and two times equivalent diameters as
edge length to draw a square box. If these square box have the
overlapping regions, they belong to one magnetic pole, otherwise
they belong to the different magnetic poles. Thus these closer
isolated regions may be one magnetic pole. But not all of the
magnetic poles could be the real poles of the BMRs in this paper.
These poles which have small area are possible not to be the real
poles of the BMRs in the active regions, because we mainly study the
larger and stronger BMRs here. Therefore, the small magnetic poles
have to be ruled out. The minimum area used here is similar to
previous studies. For example, \cite{Harvey93} studied the area of
BMRs larger than 2.5 square degrees ($\sim$373 Mm$^2$);
\cite{Tang84} had detected the ARs as small as 450 Mm$^2$; the
smallest area studied by \cite{Schrijver88} was 310 Mm$^2$;
\cite{Wang89} even studied the ARs as small as about 200 Mm$^2$.
Based on these values, we define the area (A) threshold as 250
pixel$^2$ ($\sim$495 Mm$^2$) in this paper. In other words, these
regions with the total area less than 250 pixel$^2$ are excluded as
the magnetic poles. Figure~\ref{fig1} (c) shows the positive (blue
circles) and negative (yellow circles) poles in Carrington Rotation
1960, the plus (`+') and cross (`$\times$') represent the geometric
center and the flux-weighted center, respectively. And we consider
the flux-weighted centers as the positions of the magnetic poles.
For each magnetic pole (positive or negative), we also calculate
their radius (R), fragment number (N), magnetic field (B), magnetic
flux (F), and magnetic flux density (f). Here, the magnetic pole is
supposed to be a circle. The fragment number is the amounts of the
isolated regions in the magnetic pole, and the magnetic field is
maximum value of the magnetic pole. The magnetic flux is the total
value of the magnetic pole, while the magnetic flux density is the
mean value of the magnetic filed.

\begin{figure}[h!!!]
   \centering
   \includegraphics[width=13.0cm, angle=0]{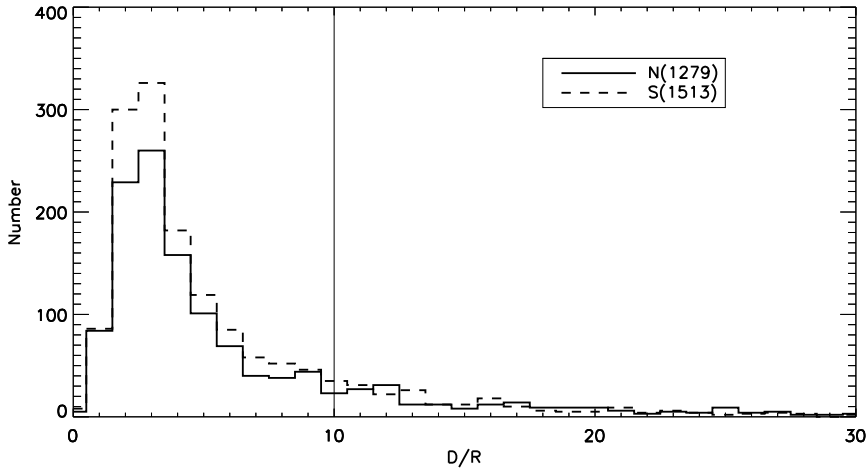}
   \caption{The ratio between the distance (D) and the radius (R) of
the magnetic poles in northern (solid line) and southern (dash line)
hemispheres, respectively.}
   \label{fig2}
\end{figure}

The second step is to identify the BMRs from the identified positive
and negative poles. In this paper, we assume that BMRs are formed by
the closest magnetic poles with opposite polarity, but the BMR
direction in northern is opposite from that in southern hemisphere.
In the northern hemisphere, a BMR is defined as a negative pole to
link the closest positive pole; while in the southern hemisphere,
the positive pole is used to link the closest negative one, such as
the red lines linking the two magnetic poles in Figure~\ref{fig1}
(c). The distance (D) between the positive and negative poles of
each BMR is measured, and the ratio of D/R is shown in
Figure~\ref{fig2}. There is a normal distribution of D/R, and the
maximum value is around 3. These BMRs with a big value of D/R, such
as greater than 10, indicating a large distance between positive and
negative poles with a small area size, are possible not real.
Therefore, these BMRs with D/R $\geq$ 10 (the vertical line in
Figure~\ref{fig2}) are ruled out to analyze in this paper, i.e., the
one marked with the rectangle. Namely, a positive pole linking with
the negative one with a shortest area-weight distance is identified
as a BMR in the southern hemisphere. And the same rule is applied in
the northern hemisphere. The red lines in Figure~\ref{fig1}~(c) link
the identified BMRs, the filled circles represent the positions of
the BMRs, which are the middle positions of the positive and
negative poles of the BMRs.

\section{Results}
\label{Res} Using the method mentioned in Section~2.2, we analyze
the SOHO/MDI data from Carrington rotation 1909 (May 1996) to 2104
(December 2010). This period covers the whole 23rd solar cycle and
the beginning phase of 24rd solar cycle. There are 4 MDI images
(CR1938,CR1939,CR1940 and CR1941) missing due to the malfunction of
the SOHO in 1998. Some MDI images (e.g., CR1937, CR1944, CR1945,
CR1956,CR2011, CR2015 and so on) are partial, but they are enough
for our analysis. Finally, 2948 positive poles and 2940 negative
poles are identified from these MDI images, and 2234 BMRs are
identified, 1005 of them are located in northern hemisphere, while
the other 1229 are located in southern hemisphere. However, there
are 3171 NOAA ARs published during the same period. Such difference
is mainly because that only these LOS magnetic fields observed near
central meridian are used. The criterion to identify the BMRs and
NOAA ARs is also different.

\subsection{The positions of BMRs}
\begin{figure}[h!!!]
   \centering
   \includegraphics[width=13.0cm, angle=0]{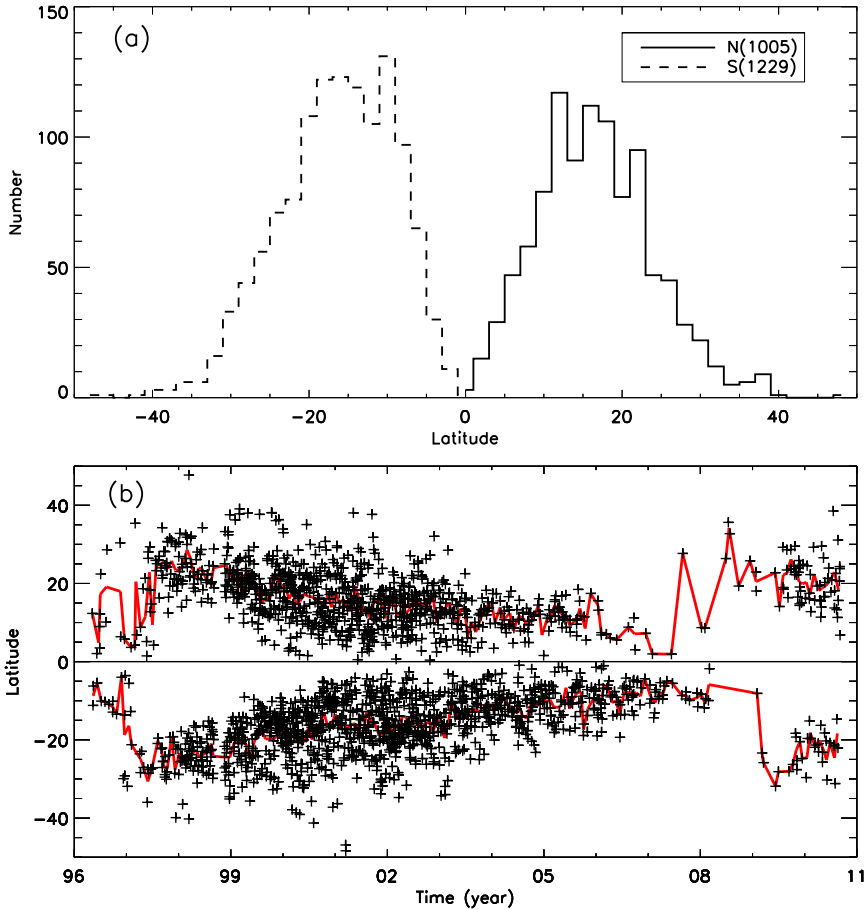}
   \caption{(a): The distribution of BMR latitudes in northern (solid
line) and southern (dash line) hemispheres, respectively. (b): The
distribution of BMR latitudes with solar cycles, and the two red
lines show the flux-weighted centers of BMRs per Carrington rotation
for the two solar hemispheres.}
   \label{fig3}
\end{figure}

In our study, each BMR has its own latitude and time. On the MDI
synoptic image, the midpoint of the positive and negative poles is
marked as the BMR position, which is the function of the latitude
and time. Figure~\ref{fig3} (a) shows the distribution of the BMR
latitudes. Most (95.7\%, 2138/2234) of the BMRs are located in the
low latitude in the solar disk, i.e., between -30$^{\circ}$ and
30$^{\circ}$; while only 4.3\% (96/2234) BMRs are exceed this range
but not higher than $\pm$50$^{\circ}$. Here, `+' and `-' represent
the latitudes of the BMRs which located in the northern and southern
hemispheres, respectively. This is consistent well with the ARs and
sunspots, which are also typically located at the low latitude in
the solar surface. The mean values of the BMR latitudes are
16$^{\circ}$ and -16$^{\circ}$ in northern and southern hemispheres,
respectively. Panel (b) further gives the BMR latitudes in solar
disk varying with solar cycles, it appears similar as the butterfly
diagram described by \cite{Maunder04,Maunder22}, which explained as
the emergence positions of the ARs (or sunspots) progressively
drifting toward the solar equator \citep{Hathaway03}. And this
further confirms that the BMRs in this paper are essentially bipolar
fields.

\subsection{The parameters of the BMRs}
\begin{figure}[h!!!]
   \centering
   \includegraphics[width=13.0cm, angle=0]{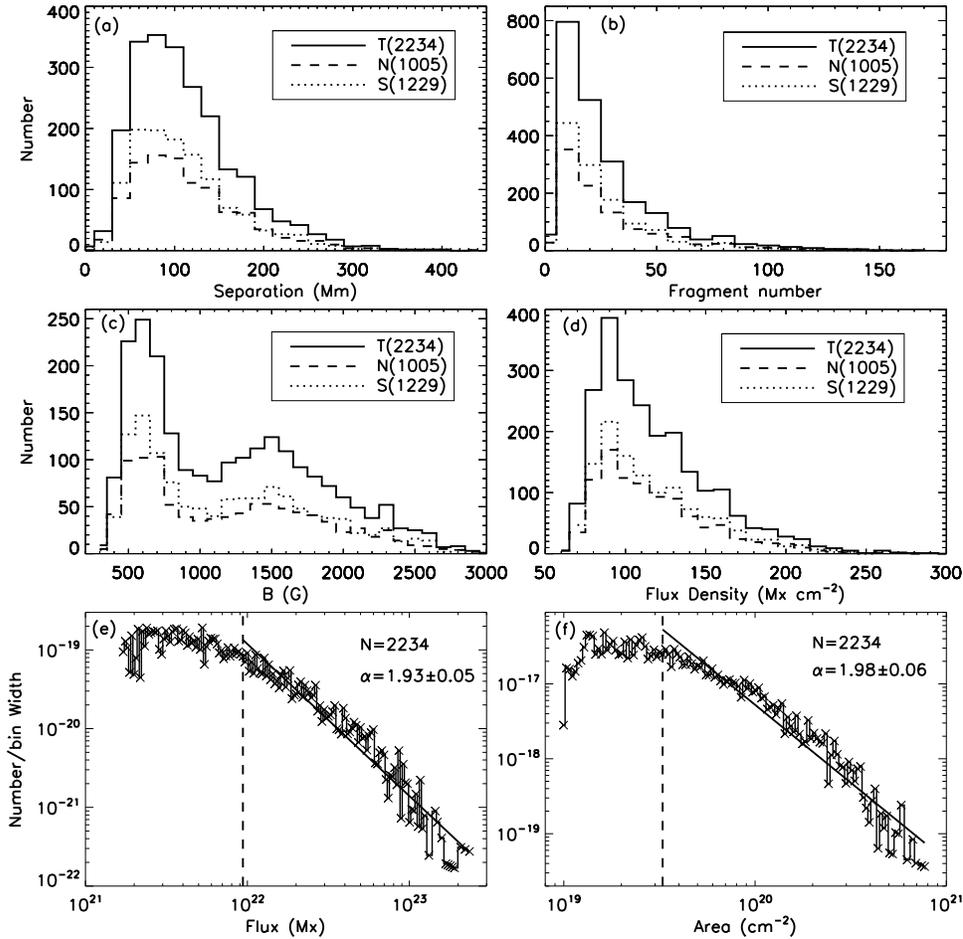}
   \caption{The solid profiles are the distributions of the BMR separated
distances (a), fragment number (b), magnetic field (c), magnetic
flux density (d). The dashed and dotted profiles represent the BMR
parameter distributions in northern and southern hemispheres,
respectively. The other two panels are the distribution of the BMR
flux (e) and area (f) in log-log space.}
   \label{fig4}
\end{figure}

\begin{figure}[h!!!]
   \centering
   \includegraphics[width=13.0cm, angle=0]{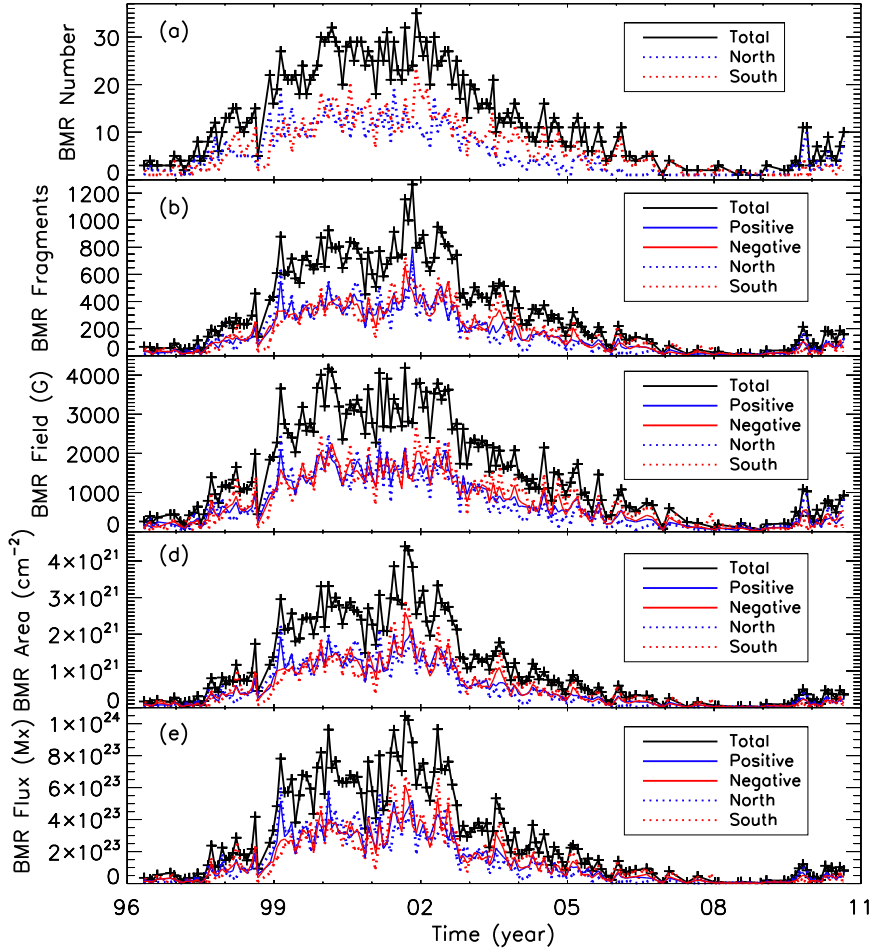}
   \caption{The BMR parameters per Carrington rotation with time from
1996 to 2010. (a): the BMR number (black); (b): the BMR total
fragment number (black), positive fragment number (blue), and
negative fragment number (red); (c): the BMR total magnetic field
(black), positive magnetic field (blue), and negative magnetic field
(red); (d): the BMR total area (black), positive area (blue), and
negative area (red); (e): the BMR total flux (black), positive flux
(blue), and negative flux (red). The dotted profile shows the
parameters in northern (blue dashed) and southern (red dashed)
hemispheres.}
   \label{fig5}
\end{figure}

For these identified BMRs, we statistical study their separations
(D$_S$), fragment number (N), area (A), magnetic filed (B), magnetic
flux (F), magnetic flux density (f). The separations are the
distance between the positive and negative poles of the BMRs, and
the fragment number, area and magnetic flux are the sum of absolute
values for positive and negative poles of the BMRs, while the
magnetic field and magnetic flux density are the half of the total
absolute values for positive and negative poles of the BMRs. The
statistical results are shown in Figure~\ref{fig4} and~\ref{fig5}.
While table~\ref{tab1} and~\ref{tab2} also list these parameters in
the documents to compare with the previous findings. Here AR1 and
AR2 are the active regions which identified from the different
definitions, i.e., AR1 is the active region identified by
\cite{Zhang10} based on their automated method, while AR2 is the
active region defined from the NOAA manual catalog.

\begin{table}
\bc
\begin{minipage}[]{100mm}
\caption[]{Statistical results of the BMRs and ARs.} \label{tab1}
\end{minipage}
\begin{tabular}{lcccccccccccccc}
  \hline\noalign{\smallskip}
Parameter  & \multicolumn{3}{c}{Mean} & \multicolumn{3}{c}{Median} & \multicolumn{3}{c}{Min}  &  \multicolumn{3}{c}{Max}  \\
           & BMR & AR1 & AR2  & BMR & AR1 & AR2  & BMR & AR1 & AR2 & BMR  & $^1$AR1  & $^2$AR2    \\

  \hline
D$_S$ (Mm)           & 112 & - & - & 100 & - & - & 3.2 & -  & - & 448 & -  & -   \\
N                    & 26  & 7.8 & 8 & 19  & 5 & 4 & 2   & 1  & 0 & 168 & 69  & 90   \\
A ($10^{19}$ cm$^2$)  & 8.8 & 6.1 & 0.3 & 5.5 & 3.5 & 0.06 & 1.0 & 0.3  & 0 & 79.7 & 68  & 6.9   \\
B (G)                & 1194 & - & - & 1107 & - & - & 252 & -  & - & 2972 & -  & -   \\
f (Mx cm$^2$)        & 117 & - & - & 109 & - & - & 58 & -  & - & 339 & -  & -   \\
F ($10^{21}$ Mx)      & 22.4 & 16.7 & - & 12.3 & 8.4 & - & 1.7 & 0.9  & - & 243 & 197 & -   \\
  \noalign{\smallskip}\hline
\end{tabular}
\ec \tablecomments{1.0\textwidth}{$^1$AR1: the parameters of active
regions are cited from \cite{Zhang10}. \\ $^2$AR2: the parameters of
active regions are from the NOAA manual catalog.}
\end{table}

Figure~\ref{fig4} shows the distribution of each BMR parameter, and
their typical values are listed in table~\ref{tab1}. The separated
distances (D$_S$) range from 3.2 Mm to 448 Mm, which is shown in
panel~(a). The largest separation in this paper is greater than the
value of 26 deg ($\sim$320 Mm), which obtained by \cite{Howard89}.
But the number of such large separation is very rare, for example,
only 1 separation of the BMR exceeds 400 Mm, while only 0.9\%
(20/2234) separation of the BMRs exceed 300 Mm. The mean separation
distance of these BMRs is 112 Mm, which is similar to the value of
86 Mm \citep{Howard89}. The fragment number (N) of these BMRs are
from 2 to 168, and the average number of them are about 26, which is
shown in panel~(b). From panel~(a) and (b), we can see that the
distribution is similar to that of the separation distances, the
larger number is very less, only 1.9\% (42/2234) are greater than
100. The same results are applied to the distribution of the BMR
flux density, as shown in panel~(d). The minimum flux density is 58
Mx cm$^{-2}$, the maximum flux density is 339 Mx cm$^{-2}$ and the
average flux density is 117 Mx cm$^{-2}$, while only 0.45\%
(10/2234) of them exceed 250 Mx cm$^{-2}$. However, the distribution
of the BMR magnetic filed is different. As shown in panel~(c), there
are two peaks in the distribution of the magnetic field, one is
about 600 G, the other is around 1500 G, and the magnetic strength
ranges from 252 G to 2972 G.

For these 2234 BMRs, their area and magnetic flux are also measured.
The area of these BMRs ranges from 1.0$\times$10$^{19}$ cm$^2$
($\sim$1000 Mm$^2$) to 7.97$\times$10$^{20}$ cm$^2$
($\sim$7.97$\times$10$^{4}$ Mm$^2$). The smallest area of BMRs in
this paper is larger than previous results of 300$-$400 Mm$^2$ for
the ARs \citep{Tang84,Schrijver88,Harvey93,Zhang10}, and the largest
area is also greater than the earlier results of about
(1$-$7)$\times$10$^{4}$ Mm$^2$ for the ARs
\citep{Tang84,Schrijver88,Wang89,Harvey93,Zhang10}. The flux of
these BMRs are from 1.7$\times$10$^{21}$ Mx to 2.43$\times$10$^{23}$
Mx, which is similar to the AR flux from 8.6$\times$10$^{20}$ Mx to
1.97$\times$10$^{23}$ Mx obtained by \cite{Zhang10}, but the maximum
value is larger than earlier results of about 10$^{22}$ Mx for ARs
\citep{Howard89,Wang89}. Then the frequency distributions of these
BMRs occurrence as functions of flux and area are shown in the two
bottom panels of Figure~\ref{fig4}. It is hard to say that the
behaviors of the frequency distributions for the whole area and
flux. However, using the maximum likelihood estimation (MLE) method
developed by Clauset et al. (2009), both the BMR area and flux
exhibit a power-law behavior, i.e., $dN/dx \propto x^{-\alpha_{x}}$.
The MLE method is based on the Kolmogorov-Smirnov statistic to
determine the lower cutoff (x$_{min}$) of the power-law behavior,
which is marked by the dashed lines in panel (e) and (f). Using the
MLE method, we obtain the power-law index of $\alpha_{F}$ = 1.93
$\pm$ 0.05 for the BMR flux and $\alpha_{A}$ = 1.98 $\pm$ 0.06 for
the BMR area. This is consistent well with the power-law
distribution of the large solar activities, such as radio bursts,
soft X-rays, hard X-rays, interplanetary type III bursts,
interplanetary particle events and CMEs
\citep{Crosby93,Aschwanden98}. \cite{Dennis85} and \cite{Crosby98}
further summarize $\alpha_x$ for the distributions of different
flare-related parameters and state that it varies from 1.4 to 2.4.
This is also proved by many authors for solar flares
\citep{Wheatland00,Su06,Li12}, CMEs \citep{Wheatland03}, radio
bursts \citep{Ning07,Song12} and other small-scale magnetic fields
\citep{Parnell09,Li13}. In our results, $\alpha_{A}$ = 1.98 $\pm$
0.06 for the area of the BMRs are followed by the fractal models
\citep{Aschwanden02}. While $\alpha_{F}$ = 1.93 $\pm$ 0.05 for the
BMR flux are consistent with the index of coronal activities (i.e.,
solar flares, CME, radio burst) and bright points \citep{Li13}. Our
findings indicate that there are not fundamental differences for
their generation in solar atmosphere, whatever the (small or large)
scale or the (low or high) height above solar surface. And the
observations also show that both the coronal activities and bright
points are strongly related to the magnetic fields, indicating the
common features of the generating magnetic structures at small or
large scales.

\begin{table}
\bc
\begin{minipage}[]{100mm}
\caption[]{Statistical results of the BMRs vs ARs per Carrington
rotation.} \label{tab2}
\end{minipage}
\begin{tabular}{lcccccccccccccc}
  \hline
Parameter  & \multicolumn{3}{c}{Mean} & \multicolumn{3}{c}{Median} & \multicolumn{3}{c}{Min}  &  \multicolumn{3}{c}{Max}  \\
           & BMR & AR1 & AR2  & BMR & AR1 & AR2  & BMR & AR1 & AR2 & BMR & $^1$AR1 & $^2$AR2   \\

  \hline
N$_0$                & 11 & 11 & 13.5 & 10 & 8 & 11 & 0 & 0  & 0 & 35 & 37  & 39   \\
N                    & 312  & 85 & 126 & 210  & 55 & 101  & 0  & 0  & 0 & 1265 & 327  & 413   \\
A ($10^{20}$ cm$^2$)  & 10.2 & 6.7 & 0.5 & 5.4 & 4.3 & 0.3 & 0 & 0  & 0 & 44.0 & 26.7  & 1.8   \\
B ($10^{3}$ G)        & 1.39 & - & - & 1.07 & - & - & 0 & 0  & 0 & 4.19 & -  & -   \\
f ($10^{3}$ Mx cm$^2$) & 1.37 & - & - & 1.13 & - & - & 0 & 0 & 0 & 4.19 & -  & -   \\
F ($10^{23}$ Mx)      & 2.6 & 1.8 & - & 1.4 & 1.3 & - & 0 & 0  & - & 10.5 & 6.96  & -   \\
  \hline
\end{tabular}
\ec \tablecomments{1.0\textwidth}{$^1$AR1: the parameters of active
regions are cited from \cite{Zhang10}. \\ $^2$AR2: the parameters of
active regions are from the NOAA manual catalog.}
\end{table}

Figure~\ref{fig5} displays the variation of BMR parameters per
carrington rotation with the period from 1996 to 2010. The typical
values of these BMR parameters per Carrington rotation are listed in
table~\ref{tab2}. Here N$_0$ is the number of the BMRs or ARs per
carrington rotation, the other symbols is same as in
table~\ref{tab1}. As shown in Figure~\ref{fig5}, there are double
peaks during solar maximum. And if we examine carefully the panels
(c), (d) and (e), we could find that the second peak flux (in late
2001) of BMRs are mainly caused by the large area of the emerged
BMRs, but not the mean strength of the magnetic field. Based on
this, the 23rd solar cycle is peaked in late 2001 but not in early
2000. These results are consistent well with previous findings
\citep{Zhang10}. From table~\ref{tab2} we can see that the BMR
parameters are consistent well with the AR parameters except for the
fragment number. In Figure~\ref{fig5}, we also plot the BMR
parameters vary with solar cycles in the northern (blue dashed line)
and southern (red dashed line) hemispheres, which clearly show the
north-south asymmetry of the BMR distribution. And this north-south
asymmetry is also shown in panels (a) $-$ (d) of Figure~\ref{fig4}.
Similar north-south asymmetry of ARs has been reported earlier
\citep{Temmer02,Zharkov06,Zhang10}. However, this asymmetry is not
well understood yet. We also plot the parameters of positive (blue
solid line) and negative (red solid line) poles vary with solar
cycles, and the positive and negative parameters of the BMRs are
almost the same during the solar cycles, although there may be some
difference in one or two carrington rotation periods. In
table~\ref{tab2}, the parameters of the BMRs and ARs are zero in
solar minimum, whether in our data or in other data. This is because
that the chance of appearing ARs is very small or even not in one
carrington rotation during solar minimum.

\subsection{The orientation of BMRs}

\begin{figure}[h!!!]
   \centering
   \includegraphics[width=10.0cm, angle=0]{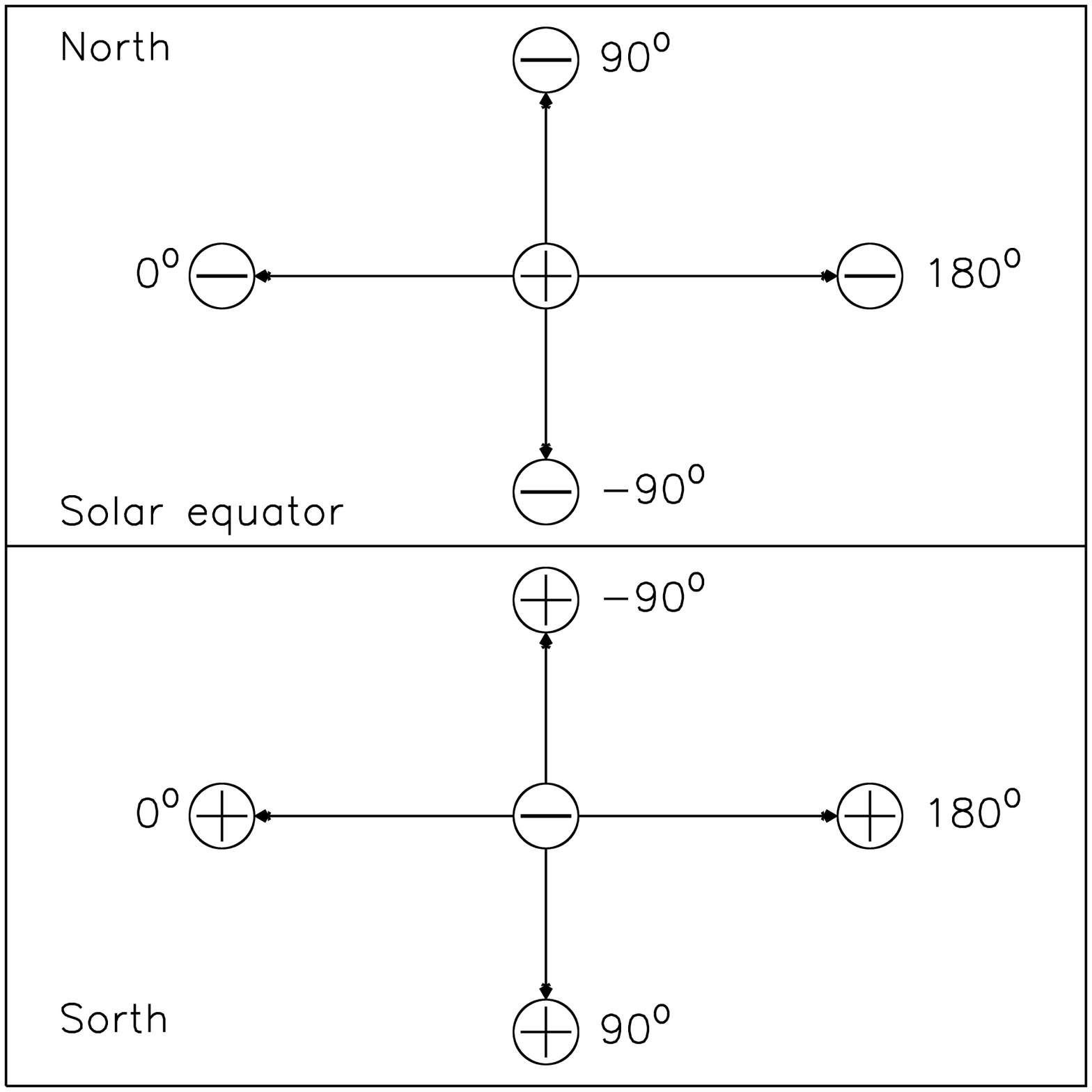}
   \caption{Schematic representation of the orientation angle
($\theta$) of the BMRs. In northern hemisphere the positive pole
(`+') is the angle center, and the negative pole (`-') rotation
clockwise; while in southern hemisphere, the negative pole is the
angle center, and the positive pole rotation counter-clockwise. The
angle ranges from -90$^{\circ}$ to 270$^{\circ}$. }
   \label{fig6}
\end{figure}

The orientation angles of the ARs are important to understand the
solar cycles and have been studied in the past
\cite[e.g.,][]{Wang89,Howard89,Howard91a}. However, most of the data
they used are fully-disk magnetograms which observed on ground. In
this paper, we use the MDI carrington rotation charts to study the
orientation angles. Figure~\ref{fig6} gives the definition of the
orientation angle ($\theta$) of BMRs in northern and southern
hemispheres, respectively. In northern hemisphere, the orientation
angle ($\theta$) is defined to be the angle of vector originating in
the flux-weighted center of the positive pole and rotating clockwise
and terminating in the flux-weighted center of the negative pole;
while it is defined to be the angle of vector originating in the
flux-weighted center of the negative pole and rotating
anti-clockwise and terminating in the flux-weighted center of the
positive pole in southern hemisphere. Noting that we separately
define the BMR orientation angle in northern and southern
hemisphere. Both the BMR orientation angles in northern and southern
hemispheres range from -90$^{\circ}$ to 270$^{\circ}$. In this
paper, the normally oriented BMRs refer to those BMRs whose
orientation angles range from -90$^{\circ}$ to 90$^{\circ}$, while
the abnormal orientation angles are between 90$^{\circ}$ and
270$^{\circ}$ in solar disk, whatever in northern hemisphere or in
southern hemisphere.

\begin{figure}[h!!!]
   \centering
   \includegraphics[width=15.0cm, angle=0]{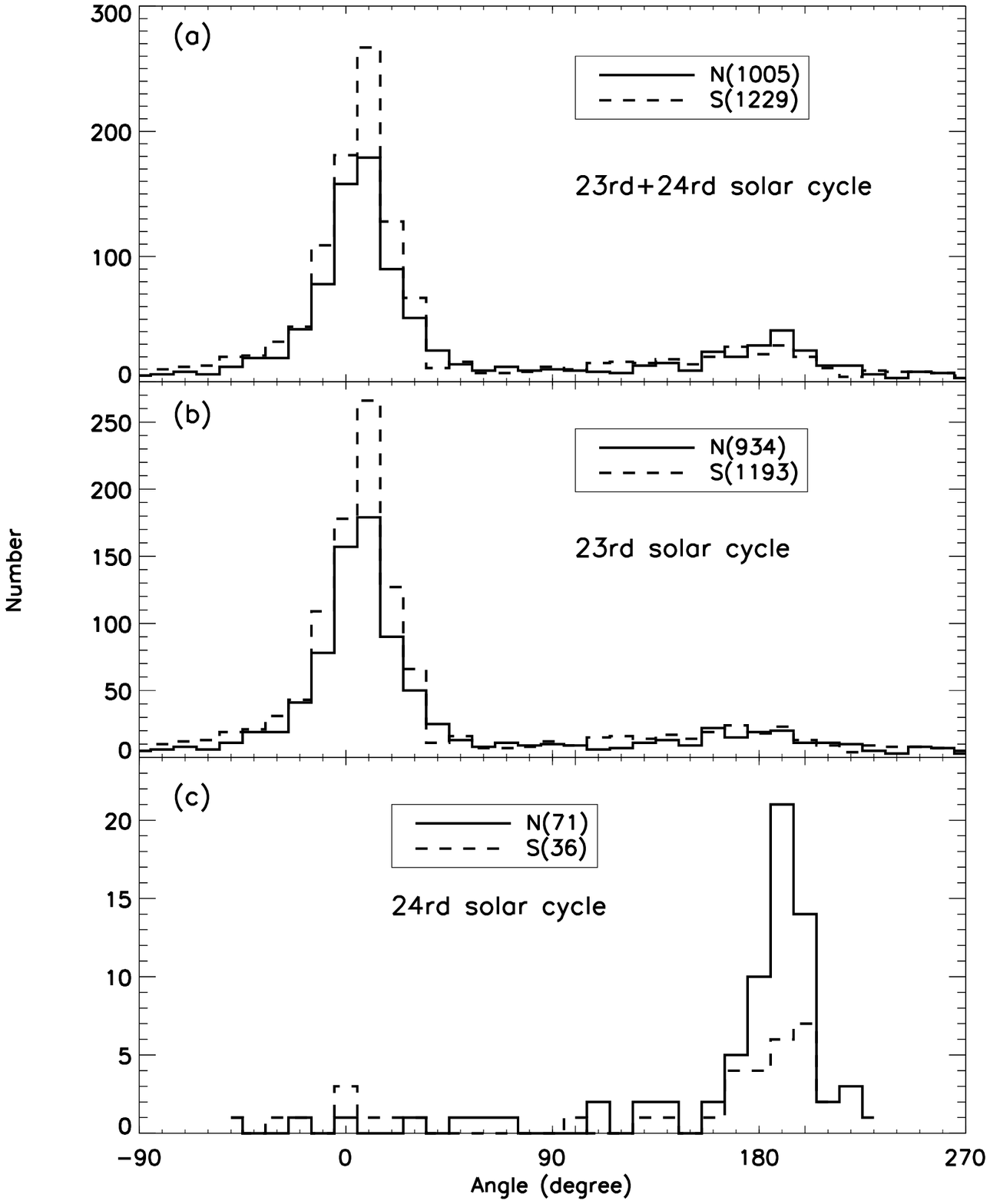}
   \caption{The BMR orientation angle ($\theta$) distribution with a
bin of 10$^{\circ}$ from the data during the 23rd and 24rd solar
cycle (a), from the data during the 23rd solar cycle (b) and from
the data during the beginning of 24rd solar cycle (c). While the
solid and dashed lines represent the BMRs in northern and southern
hemispheres, respectively.}
   \label{fig7}
\end{figure}

Figure~\ref{fig7} gives the orientation angle ($\theta$)
distribution of these BMRs with a bin of 10$^{\circ}$ in northern
(solid lines) and southern (dashed lines) hemispheres, respectively.
Panel (a) shows all the BRMs are from carrington rotation 1909 to
2104, including the whole 23rd solar cycle and the beginning phase
of 24rd solar cycle. Since the orientation angles are ranging from
-90$^{\circ}$ to 270$^{\circ}$ and most of them are focus on around
0$^{\circ}$ or 180$^{\circ}$. In this paper, the median values of
the whole data are 13$^{\circ}$ and 11$^{\circ}$ in northern and
southern hemispheres, respectively, indicating that whether in
northern hemisphere or southern hemisphere, the orientation angles
of the BMRs are slightly deviated to the solar equator direction. In
panel~(a), we can see that there are two peaks in both northern and
southern hemispheres. And there are much more BMRs around
0$^{\circ}$ than that around 180$^{\circ}$. Then we separated the
BMRs into the entire 23rd solar cycle (from CRs 1909 to 2070) and
the beginning phase of 24rd solar cycle (from CRs 2071 to 2104), as
shown in panels~(b) and (c). From which, there is only one peak
whether in northern or southern hemispheres at the entire 23rd solar
cycle, and the median values of the BMR orientation angles are
11$^{\circ}$ and 10$^{\circ}$ in northern and southern hemispheres,
which is similar to that of all the data. And at the beginning of
24rd solar cycle, there is also one peak both in northern and
southern hemispheres, but the median values of the BMR orientation
angles are 187$^{\circ}$ and 181$^{\circ}$ in northern and southern
hemispheres, respectively. These results suggest that the two peaks
of the orientation angles in panel~(a) are from the 23rd and 24rd
solar cycles, respectively.

\begin{figure}[h!!!]
   \centering
   \includegraphics[width=13.0cm, angle=0]{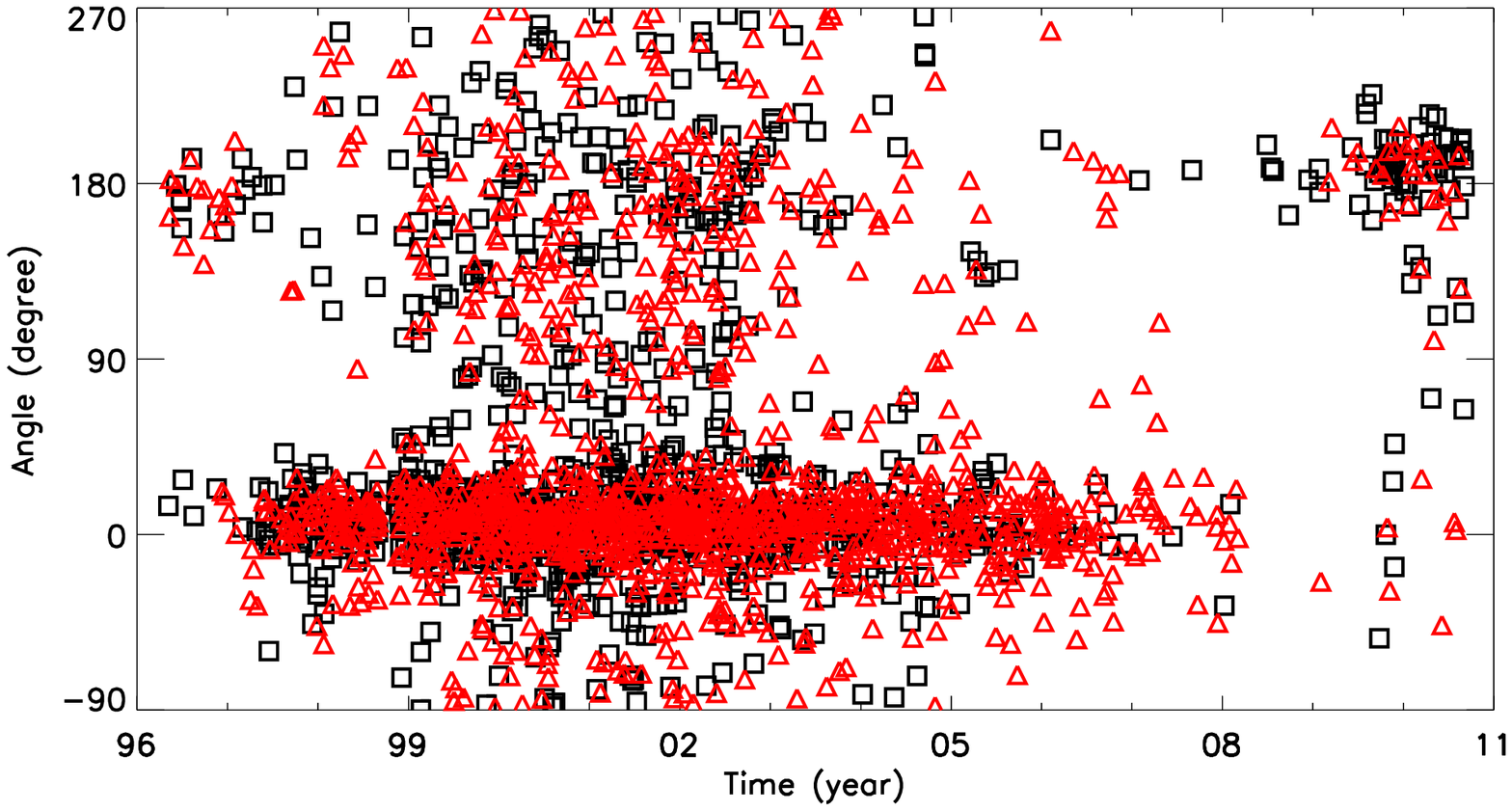}
   \caption{The time evolution of the BMR orientation angles ($\theta$)
from May 1996 to December 2010 in northern (squares) and southern
(triangles) hemispheres.}
   \label{fig8}
\end{figure}

Our results show that the orientation angles of the BMRs are
generally slightly deviated to the solar equator direction (panel
(a) in Figure~\ref{fig7}), and that the directions of the BMRs are
opposite in northern and southern hemispheres (panel (a) in
Figure~\ref{fig7}), and that the magnetic polarity of BMRs is
reversed in solar minimum of the solar cycle (panel (b) and (c) in
Figure~\ref{fig7}). These results are also shown in
Figure~\ref{fig8}, which gives the variation of BMR orientation
angles with respect to the phase of solar cycles from May 1996 to
December 2010. The squares and triangles represent the BMRs located
in the northern and southern hemispheres, respectively. In this
figure, the BMR orientation angles vary with solar cycles, and the
polarities of BMRs are changing in different solar cycles. We note
that the orientation angles are reversed around in June 2008. That
is to say, the 23rd solar cycle is from 1996 to 2008, and the
duration is about 12 years. This perhaps can be explained by the
extended activity cycle, because the sunspot region of a new solar
cycle could begin to appear as much as $\sim$1.6 years before the
defined solar minimum and continue to emerge up to $\sim$1.8 years
after the following minimum. These results are consistent well with
the previous findings about the ARs
\citep{Howard89,Wang89,Howard91a} and sunspots
\citep{Maunder04,Hale19,Howard91b}, indicating that all the
large-scale phenomenons related to the bipolar magnetic fields are
following by the ``Hale's Polarity Laws''. That is to say, during
the first solar cycle the positive poles of the BMRs are the leading
polarity in northern hemisphere and the negative poles of the BMRs
are the leading polarity in southern hemisphere; while in the next
solar cycle, the BMR polarity is reversed, the leading polarity are
negative poles in northern hemisphere and positive poles in southern
hemisphere.

\begin{figure}[h!!!]
   \centering
   \includegraphics[width=13.0cm, angle=0]{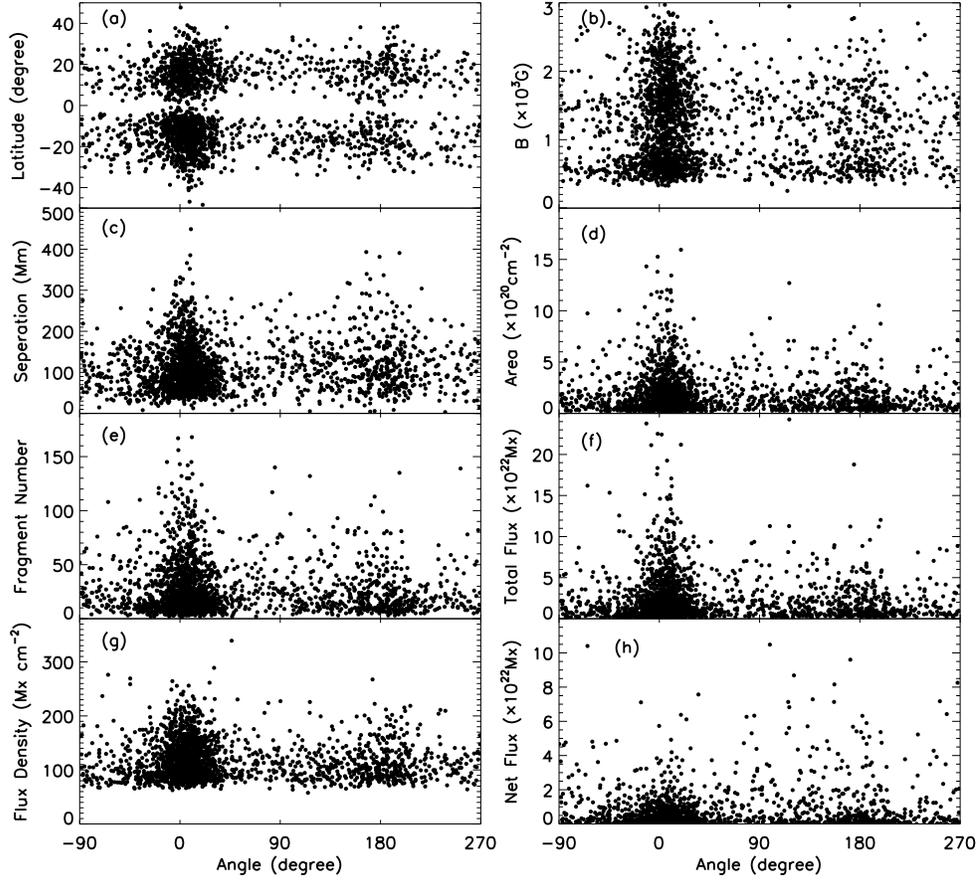}
   \caption{The BMR orientation angles ($\theta$) dependence on their
latitudes~(a), magnetic filed~(b), separations~(c), area~(d),
fragment number~(e), total flux~(f), flux density~(g) and net
flux~(h).}
   \label{fig9}
\end{figure}

\begin{figure}[h!!!]
   \centering
   \includegraphics[width=13.0cm, angle=0]{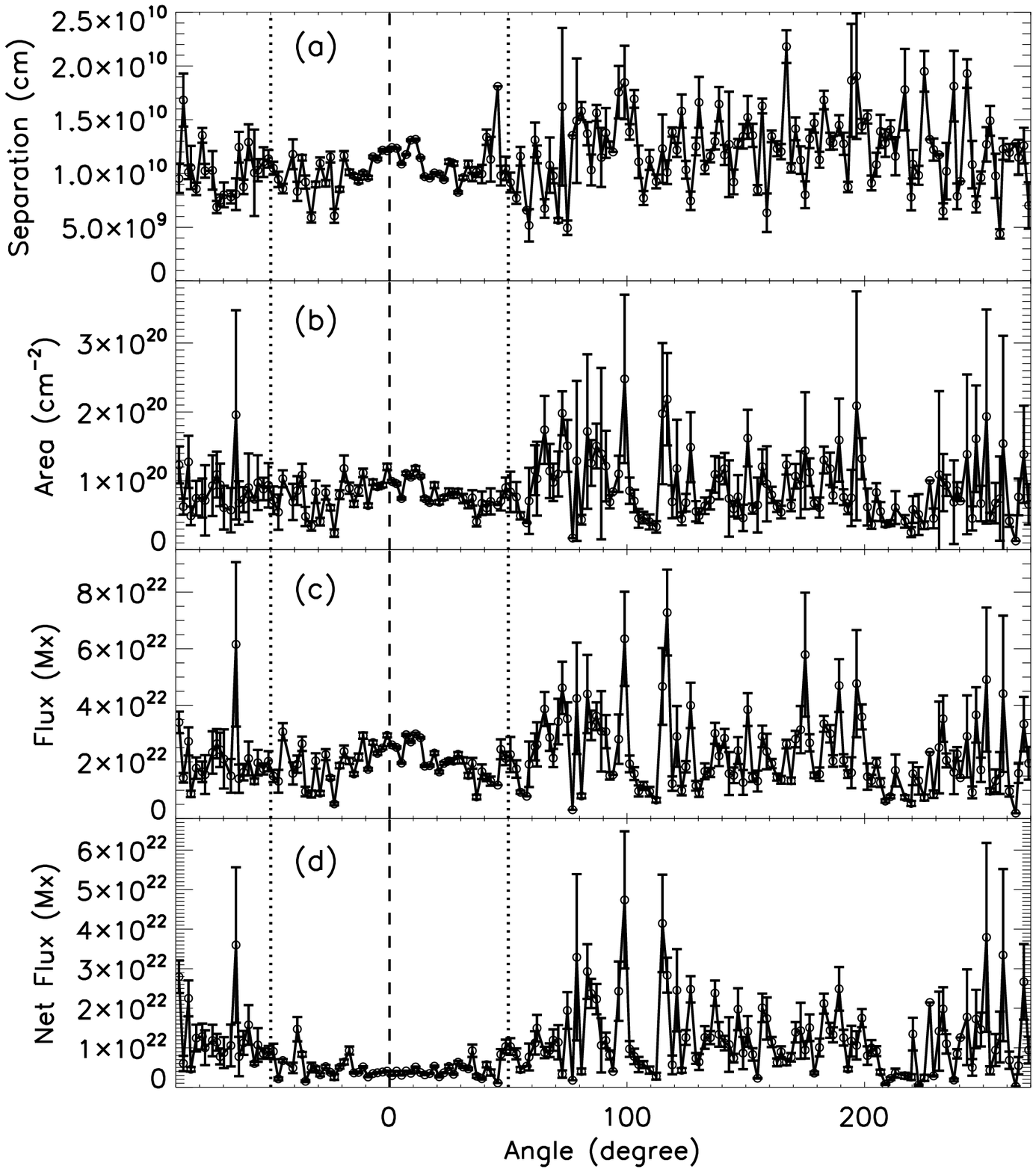}
   \caption{The average BMR parameters (separations~(a), area
sizes~(b), total magnetic flux~(c) and net magnetic flux~(d)) for 2
degree intervals of BMR orientation angles. The error bars represent
the standard deviations of those mean values, and the dashed lines
are corresponding these mean values with 0 degree, while the dotted
lines corresponding these mean values with -50 or 50 degree.}
   \label{fig10}
\end{figure}

Figure~\ref{fig9} displays the BMR orientation angles dependence on
their latitudes~(a), magnetic filed~(b), separations~(c), area~(d),
fragment number~(e), total flux~(f) (the sum values between the
absolute positive and negative flux), flux density~(g) and net
flux~(h) (the minus values between the absolute positive and
negative flux). It is hard to see any relationship between the
orientation angles and other parameters in the BMRs, especially
between the orientation angles and their latitudes. Here all of the
2234 BMRs are analyzed, including the normal and abnormal BMRs from
May 1996 to December 2010. Then only the BMRs (2127) during the 23rd
solar cycle are selected to analyze, regardless of the northern or
southern hemispheres in solar disk. Figure~\ref{fig10} gives the
average paraments for various values of orientation angles of these
2127 BMRs. The BMRs with larger absolute orientation angles tend to
show greater deviation on average values of paraments (e.g.
separations~(a), area~(b) and flux~(c)) than those with smaller
absolute orientation angles. Especially for these normal orientated
BMRs whose orientation angles between -50$^{\circ}$ and 50$^{\circ}$
(dotted lines), their deviation and fluctuation are much smaller
than other BMRs. However, the net flux of these normal orientated
BMRs whose orientation angles between -50$^{\circ}$ and 50$^{\circ}$
(dotted lines) are always smooth and the fluctuation are very small
(panel~d). But for those other orientated BMRs (the absolute
orientation angles are exceeded 50$^{\circ}$), their fluctuation and
deviation are much larger. These results are similar to that
obtained from the ARs by \cite{Howard89,Howard91a}, who found that
these ARs with less absolute orientation angles tend to show smaller
deviation. Finally the magnetic flux of the BMRs in this paper are
larger about one magnitude than others, which possibly due to the
resolution and sensitivity of the different instruments.
\cite{Howard89} have been said that the inherent disadvantage of
their data is the poor resolution, and this is the advantage of our
data as the MDI magnetic charts have high resolution and high
sensitivity.

\begin{figure}[h!!!]
   \centering
   \includegraphics[width=13.0cm, angle=0]{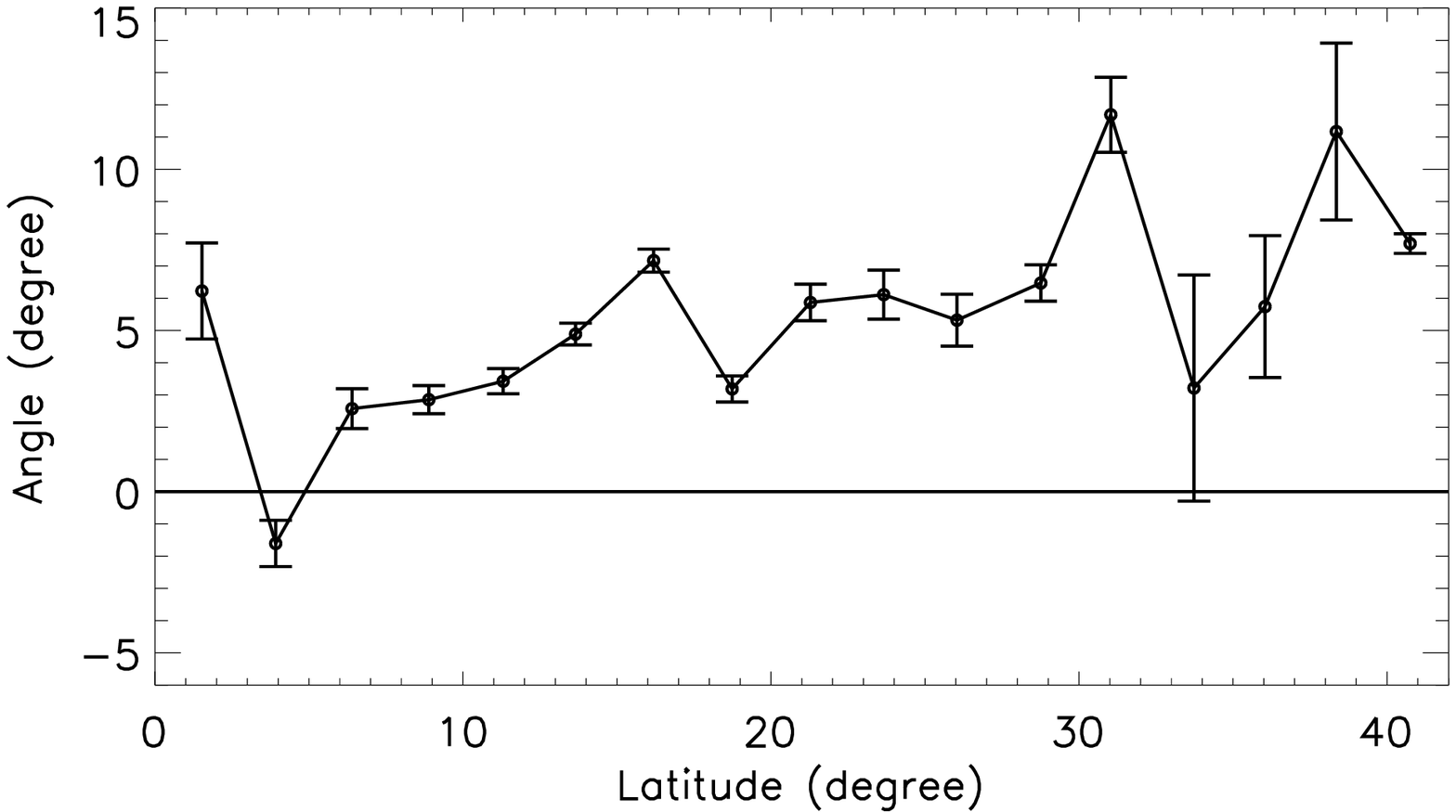}
   \caption{Average BMR orientation angles for various intervals of
absolute solar latitudes in degrees. The average values are taken
over 2.5 degree latitudes, the error bars represent the standard
deviations of those mean values.}
   \label{fig11}
\end{figure}

Based on the above analyzing, only the normal BMRs 1543 which the
orientation angles range from -50$^{\circ}$ to 50$^{\circ}$ in 23rd
solar cycle are selected to analyze, regardless of the northern or
southern hemisphere on solar disk. Figure~\ref{fig11} shows the
orientation angles dependence on the absolute latitudes for these
1543 normally oriented BMRs during the 23rd solar cycle, the average
orientation angles are taken over 2.5 degree latitudes, and the
error bars represent the standard deviations of these mean values.
For these normally oriented BMRs, the orientation angles increase
with their latitudes on the whole. This is consistent well with the
results that the tilt angles of the ARs and sunspots general
increasing with the latitudes
\citep[e.g.,][]{Hale19,Wang89,Howard89,Howard91a,Howard91b}. And at
the latitudes equator-ward of about 5 degree, the orientation angles
show the negative values on average but not very significant, which
is also similar to the results from the tilt angles of ARs
\citep{Howard91a}.

\section{Conclusions and Discussions}
\label{Con} Using SOHO/MDI LOS magnetic data from CRs 1909 to 2104,
we statistically study the observational features of BMRs. To obtain
the BMRs, we firstly identify the positive and negative poles with
the strength and area thresholds, respectively; then using the
criteria of closest area-weight distance between positive and
negative poles to determine the BMRs. Finally, 2234 BMRs are
obtained from the observation data, 1005 of them are lactated in
northern hemisphere and the other 1229 BMRs are lactated in southern
hemisphere.

For these BMRs, the average latitudes are 16$^{\circ}$ and
-16$^{\circ}$ in northern and southern hemispheres, respectively.
And most (95.7\%) of these BMRs locate between -30$^{\circ}$ and
30$^{\circ}$, while only a little (4.3\%) BMRs locate in high
latitude but not exceed 50$^{\circ}$. The time variation of these
BMR latitudes with solar cycles is similar to the classical
butterfly diagram of sunspots \citep{Maunder04,Maunder22}. These
BMRs are also follow the ``Sp\"{o}rer's Law''. The BMR separations,
area, flux are calculated (see table~\ref{tab1} and
Figure~\ref{fig4}) and are similar to the results obtained by
\cite{Zhang10}. However, the fragment number is much larger than
others, this is because that we do not limit the area size of the
fragments. The variation of BMR parameters per carrington rotation
with solar cycles is also studied in this paper, and we further
confirm that the 23rd solar cycle peaked in late 2001 but not in
early 2000, this is consistent well with the results obtained by
\cite{Zhang10}. We also find the north-south asymmetry of the BMRs,
which is similar to previous findings about the north-south
asymmetry of the ARs
\citep[e.g.,][]{Temmer02,Zharkov06,Zhang10,Shetye15}.

We find that the frequency distributions of these 2234 BMRs as
function of magnetic flux and area exhibit a power-law behavior, and
the power-law indexes, $\alpha_F$ = 1.93 $\pm$ 0.05 for the magnetic
flux, $\alpha_{A}$ = 1.98 $\pm$ 0.06 for the area, which are
consistent with the previous findings for the large-scale coronal
activities (e.g., solar flares, CMEs, radio bursts) and small-scale
magnetic elements \citep{Li13}. This suggests that the area size of
the BMRs are followed by the fractal models \citep{Aschwanden02} and
the mechanisms of the surface magnetic features on the Sun are free
of scales, indicating that all the surface magnetic features,
regardless of their scales, are generated by the similar mechanisms,
or indicating that the surface processes (i.e., fragmentation,
coalescence, cancelation and so on) in a way which lead to a
distribution of scale-free. This is also proved by \cite{Parnell09}.

Using the definition introduced in Figure~\ref{fig6}, we study the
orientation angles ($\theta$) of the BMRs. We find that most of the
BMRs display a slightly deviated to the solar equator direction on
the solar disk, whatever in the northern or southern hemispheres,
their median value are 13$^{\circ}$ and 11$^{\circ}$, respectively.
The orientation angles of these BMRs are followed by ``Hale's
Polarity Laws'', and the polarity of the BMRs is reversed in
different solar cycles. We do not find any clearly one by one
correlation between the orientation angles and other paraments for
all the BMRs. However, if only considering the average values of the
orientation angles between -50$^{\circ}$ and 50$^{\circ}$ for the
normally oriented BMRs during the 23rd solar cycle, the dependence
of the orientation angles with latitudes is found as before, which
is that the orientation angles increase with the latitudes (see.
Figure~\ref{fig11}). But for all the BMRs, these with less absolute
orientation angles tend to show smaller deviation
\citep[e.g.,][]{Hale19,Wang89,Howard89,Howard91a,Howard91b}.

\begin{figure}[h!!!]
   \centering
   \includegraphics[width=13.0cm, angle=0]{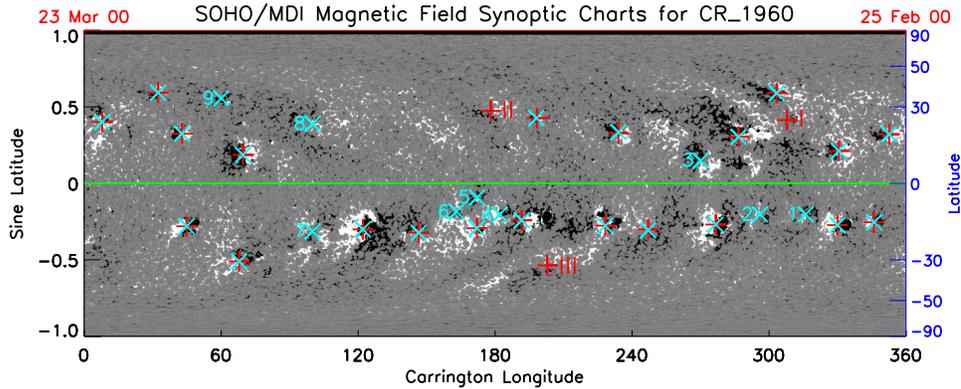}
   \caption{The synoptic charts from SOHO/MDI magnetograms for CR
   1960. The red pluses (`+') mark the positions of the identified
   BMRs, and the turquoise crosses (`$\times$') indicate the
   sites of the the NOAA ARs.}
   \label{fig12}
\end{figure}

In this paper, the BMRs in active regions are defined by the closest
magnetic poles with opposite polarities, which could result into
some wrongly or missing BMRs. Therefore, the errors estimation will
be given through comparing the NOAA ARs and our BMRs. Here, the NOAA
catalog is assumed to provide the `ground truth'. Fig.~\ref{fig12}
gives the MDI synoptic charts for CR 1960. In this rotation, 24 BMRs
are identified by our method, which are marked with the red pluses
(`+'). At the same time intervals and meridian regions, 30 NOAA ARs
are published, as indicted by the turquoise crosses (`$\times$').
For these identified BMRs, 70\% (21/30) of them could be found a
one-to-one correspondence with the NOAA ARs, as shown with the
overlapping symbols. For these BMRs without the corresponding NOAA
ARs, only one (`I') maybe not the real BMR because it has the same
positive pole with other BMR. The other two (`II' and `III') could
be the real BMRs as they have the larger and stronger magnetic field
with opposite polarities. Therefore, the true identified BMRs from
the NOAA ARs could be more than 70\%. On the other hand, there are 9
NOAA ARs have missed from our method. Most of these missing NOAA ARs
are possibly not the real BMRs according to our method. Because some
of these missing NOAA ARs only display one stronger poles, such as
ARs `5', `6' and `9'. Some others have two opposite but very
disperse magnetic poles, i.e., ARs `1', `2', `7', and `8'. In a
word, our method is useful for these BMRs connecting to the opposite
magnetic poles with stronger and compact magnetic fields.

Basing on the computational algorithm, we have automatically
identified 2234 BMRs in the active regions from MDI synoptic
magnetograms during 1996 to 2010. Meantime, there are total 3171
NOAA ARs published during the same intervals. We could pick up
$\sim$70.5\% (2234/3171) of the total NOAA ARs. Considering that
only these LOS magnetic fields observed near central meridian are
used in this paper, some NOAA ARs maybe not appear at the central
meridian regions. That is to say, not all the 3171 NOAA ARs could be
detected by our observations. Therefore, we can find more than
$\sim$70.5\% of the appearing NOAA ARs. On the other hand,
Fig.~\ref{fig7}~(b) shows that the peak distribution of 180 degree
is still existent but less obvious than that in panel~(a). This peak
possible duo to the wrongly identifying BMRs from our computational
algorithm. Fig.~\ref{fig8} suggests that these abnormal orientations
appear mainly in solar maximum. Based on these facts, we could pick
up the wrongly identifying BMRs in solar maximum. At last, 52 (or
77) BMRs are wrongly identified in northern (or southern)
hemisphere. And then the accuracy rate is about 94.3\%.

\normalem
\begin{acknowledgements}
The authors would like to thank the anonymous referee for his/her
valuable comments to improve the manuscript. We are indebted to the
SOHO/MDI teams and MDI Magnetic Field and Intensity Synoptic Charts
for providing the data. This work is supported by NSF of China under
Grants 11603077, 11573072, 11333009, the Youth Fund of Jiangsu No.
BK20161095, and the Laboratory No. 2010DP173032.
\end{acknowledgements}

\label{lastpage}

\end{document}